\documentclass[12pt]{article}
\usepackage[utf8]{inputenc}

\usepackage{amsmath}
\usepackage{amsfonts}
\usepackage{amssymb}
\usepackage{makeidx}
\usepackage{graphicx}
\usepackage{subcaption}
\usepackage{listings}
\usepackage{apacite}
\usepackage{natbib}
\usepackage{indentfirst}

\usepackage{siunitx}
\usepackage{multirow}
\usepackage{booktabs}

\usepackage{setspace}
\usepackage{authblk}
\usepackage[pdfborder={0 0 0}]{hyperref}

\title{Modelling Crypto Asset Price Dynamics, Optimal Crypto Portfolio, and Crypto Option Valuation}

\author[a]{Yuan Hu}
\author[b]{Svetlozar T. Rachev}
\author[c]{Frank J. Fabozzi}

\affil[a]{\small Texas Tech University\\
\url{yuan.hu@ttu.edu}}
\affil[b]{\small Texas Tech University\\
\url{zari.rachev@ttu.edu}}
\affil[c]{\small EDHEC Business School\\
\url{frank.fabozzi@edhec.edu}}
 
\begin{document}
\thispagestyle{plain}
\begin{spacing}{1.0}
\maketitle

\noindent {\textbf {Abstract}}\ \ \ \ Despite being described as a medium of exchange, cryptocurrencies do not have the typical attributes of a medium of exchange. Consequently, cryptocurrencies are more appropriately described as crypto assets. A common investment attribute shared by the more than 2,500 crypto assets is that they are highly volatile. An investor interested in reducing price volatility of a portfolio of crypto assets can do so by constructing an optimal portfolio through standard optimization techniques that minimize tail risk. Because crypto assets are not backed by any real assets, forming a hedge to reduce the risk contribution of a single crypto asset can only be done with another set of similar assets (i.e., a set of other crypto assets). A major finding of this paper is that crypto portfolios constructed via optimizations that minimize variance and Conditional Value at Risk outperform a major stock market index (the S$\&$P 500). As of this writing, options in which the underlying is a crypto asset index are not traded, one of the reasons being that the academic literature has not formulated an acceptable fair pricing model.  We offer a fair valuation model for crypto asset options based on a dynamic pricing model for the underlying crypto assets. The model was carefully backtested and therefore offers a reliable model for the underlying crypto assets in the natural world. 
We then obtain the valuation of crypto options by passing the natural world to the equivalent martingale measure via the Esscher transform.
Because of the absence of traded crypto options we could not compare the prices obtained from our valuation model to market prices.  Yet, we can claim that if such options on crypto assets are introduced, they should follow closely our theoretical prices after adjusting for market frictions and design feature nuances. 
\end{spacing}
\newpage
\section{Introduction}
\par{Cryptocurrencies are digital assets whose intended purpose is to serve the role of a medium of exchange. Although the term cryptocurrencies is the most popular term to describe these digital assets, we believe a more appropriate description is that they are crypto assets. The Bank of Israel, for example, states that bitcoin is an asset, not a currency. Thakor (2019) offers three reasons why cryptocurrencies do not exhibit many of the fundamental properties that define a currency. They are not (1) a generally accepted medium of exchange, (2) a stable store of value due to their substantial price volatility, and (3) serve as a unit of account as of this writing. Consequently, we treat cryptocurrencies as crypto assets. As of August 2019, according to CoiLore there
were 2,565 crypto assets with a total market capitalization of \$266 billion. Total trading volume was \$48 billion as of August 2019.}
\par{A stylized fact regarding crypto assets is that they exhibit extreme price volatility. For example, the mean  monthly standard deviation of the most popular crypto asset, bitcoin, has been as high as \$780.11, while the SPDR S$\&$P 500 ETF (SPY) has been \$4.05. Table 1 shows the mean monthly standard deviation and the drawdown for the four major crypto assets and the SPY over the two-year period.}
\\
   \begin{table}[h!]
	\begin{center}		
		\label{tab:Table 1}
		\caption{Mean Monthly Price Volatility of Four Major Crypto Assets and SPY Prices in 08/12/2017-07/02/2019}
			\begin{tabular}{c c c c c c}
				\toprule[1pt]
				\textrm{}&{Bitcoin}&{Ethereum}&{Litecoin}&{Bitcoin Cash}&{SPY}\\
				\hline
				\textrm{Standard Deviation(\$)}&780.11&48.74&12.50&137.24&4.05\\
				\textrm{Maximum Drawdown(\%)}&22&24&27&32&4\\
				\bottomrule[1pt]
			\end{tabular}
		\end{center}
\end{table}
\par{There are several reasons proffered for the extreme price volatility of digital currencies compared to other assets such as stocks. First, valuation is difficult. Balcilar et al (2017) showed that volume cannot help predict the volatility of bitcoin returns at any point of the conditional distribution. Unlike stocks where their intrinsic value is determined by company fundamentals, crypto assets do not have any underlying real assets. Rather, crypto values are based solely on market sentiment. Although stocks also are partially driven by market sentiment, company fundamentals provide some basis for assessing whether they are over- or under-valued. Recently, however, a model for valuing two crypto assets (bitcoin and XRP) was developed by Mitchnick and Athey (2018). Second, the market is thin. Although millennials who have concerns about governments tend to be participants, it is institutional players who are needed to supply liquidity to the crypto asset market. Yet many large institutional investors have stayed away or have had minimal participation in the crypto asset market. However, some major asset managers who are entering the nontraditional asset space are considering including crypto assets in their portfolios. Finally, there is no regulatory oversight. Consequently, certain market practices that are prohibited by securities laws in many countries such as market manipulation are not monitored in the digital currency markets. This lack of oversight can result in substantial price movements.}
\par{Our purpose in this paper is threefold. First, we identify an appropriate multivariate model for describing the return distribution of major crypto assets. Second, we construct optimal portfolios with risk budgeting based on two risk measures. Finally, we determine risk-neutral option pricing where the underlying is the optimal portfolio of crypto assets.}
\par{We have organized the paper as follows. After presenting multivariate models
with different distributions in Section 2, we then select the best model based on Monte Carlo backtesting. Approaches to constructing optimal portfolios by minimizing risk, as well as using risk
budgeting and measures of risk-adjusted return to compare different portfolios, are covered in Section 3. Risk-neutral option pricing based on the generalized hyperbolic distribution, 
specifically the normal inverse Gaussian distribution, is covered in Section 4. Section 5 concludes our paper.}
\section{Multivariate Models}
\par{We study the top seven crypto assets in this paper for the period July 25, 2017 to July 2, 2019. We use closing daily prices over the investigation period. The seven crypto assets are bitcoin (BTC), ethereum (ETH), XRP (XRP), litecoin (LTC), bitcoin cash (BCH), EOS (EOS) and binance coin (BNB). These crypto assets are chosen based on market capitalization as reported by CoinMarketCap.  In our simulation, we use as the benchmark the SPY with dividends.}
\par{We use log returns instead of prices as done in most studies. Campbell, Lo and MacKinlay (1997) provided two main features of log returns: a complete and scale-free summary of the investment opportunity and easier to handle than price series. We denote the price of the $i^{th}$ crypto asset at time $t$ as $S^{(i)}_{t}$. Given these features, as well as desirable statistical properties, such as stationarity, we transform price series $S^{(i)}_{t}$ into a log return series as follows:
\begin{equation}
	r^{(i)}_{t}=log \frac{S^{(i)}_{t}}{S^{(i)}_{t-1}},i=1,...,d,t=0,...,T
\end{equation}
where $d$ denotes number of crypto assets and $t$ is the time index.}
\par{The portfolio weight of the $i^{th}$ crypto asset at time $t$ is defined as $\omega^{(i)}_{t}$. To find the optimal model for the dynamics of return, we start with an equally weighted portfolio in this section, which leads to $\omega^{(i)}_{t}=\frac{1}{d},1 \leq i \leq d$.}
\par{The distribution of different crypto asset returns can be regarded as the marginal distribution of portfolio returns. Hence, we use marginal distributions of the portfolio returns to analyze the dynamics for each crypto asset and correlation between crypto assets.}
\subsection{Multivariate ARMA-GARCH Models}
\par{A common model used in working with financial time series is the ARMA-GARCH model, which is the mixed autoregressive moving-average (ARMA) model of Whittle (1951) and the generalized autoregressive conditional heteroscedastic (GARCH) model of Bollerslev (1986).}
\par{Specifically, we use ARMA(1,1)-GARCH(1,1) to model returns for each crypto asset:
\begin{align}
	r^{(i)}_{t}&=\mu^{(i)}_{t}+a^{(i)}_{t},i=1,...,d,t=0,...,T
\end{align}
The drift $\mu^{(i)}_{t}$ is modeled by ARMA(1,1):	
\begin{align}
	\mu^{(i)}_{t}&=\phi^{(i)}_{0}+\phi^{(i)}_{1}r^{(i)}_{t-1}+\theta^{(i)}_{1}a^{(i)}_{t-1}
\end{align}
and the volatility $\sigma^{(i)}_{t}$ is modeled by GARCH(1,1):
\begin{align}
	a^{(i)}_{t}&=\sigma^{(i)}_{t}\epsilon^{(i)}_{t}
	\\
	(\sigma^{(i)}_{t})^2&=\alpha^{(i)}_{0}+\alpha^{(i)}_{1}(a^{(i)}_{t-1})^2+\beta^{(i)}_{1}(\sigma^{(i)}_{t-1})^2
\end{align}
where $\epsilon^{(i)}_{t}$ is the sample innovation with arbitrary distribution with zero mean and unit variance.}
\par{After calibrating the coefficients under this structure, the uncertainty is the sample innovation. Instead of studying crypto asset returns, we study their sample innovations. The main task becomes finding an appropriate distribution of the sample innovations for each crypto asset and the best joint distribution between the sample innovations for multivariate modeling.}
\par{We start by assuming that the distribution of each sample innovation is Gaussian.\footnote{When ARMA(1,1)-GARCH(1,1) model with Gaussian innovations is fitted to a time series, and the sample innovations are non-Gaussian but still have finite fourth moment, the parameters estimated are still asymptotically unbiased. However, their confidence bounds increase. Bootstrapped confidence bounds can be derived, but those are beyond the scope of this paper, as we have performed extensive backtesting to show that our model has satisfactory predictive power. We do not search for the best time series model. Our aim is to offer a relatively simple multivariate time series model with satisfactory out-of-sample performance. Two of the authors  of this paper have spent more than 30 years in the real financial world, and are well aware that the search for the “best model” for describing financial phenomena is the worst enemy of a “sufficiently good model”.} However, after checking QQ plots and alpha stable analysis (see Borak, 2005), we add the Student's {\textit t} distribution as an innovation distribution. 
We then approach the joint distribution of sample innovations in three ways: (1) {\textit t} copula, (2) multivariate {\textit t} distribution, and (3) multivariate variance-gamma distribution by Wang (2009) as explained below.
\begin{itemize}
	\item [1.]
	The {\textit t} copula is given by (see Demarta and McNeil, 2005):
	\begin{equation}
		c^t_{\nu,\Sigma}=\frac{f_{\nu,\Sigma}(t^{-1}_{\nu}(u_1),...,t^{-1}_{\nu}(u_d))}{f_{\nu}(t^{-1}_{\nu}	(u_1))...f_{\nu}(t^{-1}_{\nu}(u_d))},u\in (0,1)^d
	\end{equation}
	We use the kernel density function to transform the sample innovations into unit space in order to calculate the {\textit t} copula. In our simulation, we specify the width of smoothing window as $w_s$. 
	\item [2.]
	The probability density function of multivariate {\textit t} distribution $t_d (\nu,\mu,\Sigma)$ is given by
	\begin{equation}
		f(x)=\frac{\Gamma(\frac{\nu+d}{2})}{\Gamma(\frac{\nu}{2})\sqrt{(\pi \nu)^d |\Sigma|}}(1+\frac{(x-\mu)'\Sigma^{-1}(x-\mu)}{\nu})^{-\frac{\nu+d}{2}}
	\end{equation}
	We adjust the degrees of freedom $\nu$ to find the best fit. 
	To limit the heaviness of the sample innovation distribution tail, we set the degrees of freedom to be greater 	than four in order to have finite kurtosis.
	\item [3.]
	From Wang (2009) and Hitaj (2013), a random vector $X$ follows the multivariate variance-gamma (MVG) model if each component $X_{i}$ can be expressed as:
	\begin{align}
		X_{i}&=\mu _{i0}+A_{i}+Y_{i},i=1,...,d
		\\
		A_{i}&=\theta_{i} V+\sqrt{V}D_i
		\\
		Y_{i}&=\theta_{i} G_{i}+\sigma_{G_i}\sqrt{G_i}W^Y_i
	\end{align}
where $Y_i,Y_j$ and $A_i$ are independent for $1 \leq i,j\leq d\ (i\neq j)$. In this model, $V$ is the dependent part of the MVG distribution and $G_i$ is the independent part of each VG distribution. As suggested by Wang (2009), we fix the degrees of freedom of the dependent part as $\nu_0=min(\nu_i)$.
\end{itemize}}
\par{Finally, we generate one-step forecasts for sample innovation. Combining calibrated coefficients and sample innovation forecasts, crypto asset return forecasts are available. We generate 10,000 scenarios for each crypto asset in each roll.}
\subsection{Backtesting}
\par{To minimize portfolio risk, we use two widely used measures of market risk, Value at Risk (VaR) and Conditional Value at Risk (CVaR). Our backtesting is based on different distributions at the $1-\alpha$ confidence level ($\alpha=0.01$) for VaR and CVaR. Let $F(x)=Pr\{r\leq x\}$ denote the cumulative distribution function of crypto asset return $r$. The VaR and CVaR are defined as}
\begin{align}
	VaR_{\alpha}(x)&=-inf\{x \mid F(x)> \alpha , x\in \mathbb{R}\}\\
	CVaR_{\alpha}&=\frac{1}{\alpha} \int_{0}^{\alpha}VaR_{\gamma}(x)d\gamma
\end{align}
\par{The backtesting is simulated using Monte Carlo method using an equally weighted portfolio. The dataset consists of 707 daily returns of the top seven crypto assets in the period of 07/25/2017 to 07/02/2019, partitioned into an in-sample data from 07/25/2017 to 04/03/2018 and an out-of-sample data from 04/04/2018 to 07/02/2019.}
\par{From the backtesting results reported in Table 2, the optimal multivariate model is ARMA(1,1)-GARCH(1,1) (Gaussian innovation) with multivariate {\textit t} distribution with five degrees of freedom. The corresponding backtesting plot using Monte Carlo simulation is shown in Figure 1. In this case, the number of failures and the ratio of failures are minimized. The traffic light and binomial tests show the correctness of the model based on the binomial distribution. Hence, in the remainder of this paper we use this model to build crypto asset portfolios and derive an option price.}
\begin{table}[h]
	\begin{center}
		\label{tab:Table 2}
		\caption{Backtesting Results from 04/04/2018 to 07/02/2019}
			\begin{tabular}{l l l l l l l l }
				\toprule[1.2pt]
				\textrm{Innovation Distribution} & \multicolumn{4}{c}{\textrm{Gaussian}} & \multicolumn{3}{c}{\textrm{Student's {\textit t}}}\\
				\cmidrule(r){2-5} \cmidrule(r){6-8}
				\textrm{Joint Distribution} & \multicolumn{3}{c}{\textrm{Multi {\textit t}}}& \textrm{MVG} & \multicolumn{3}{c}{\textrm{{\textit t} copula}}\\
				\cmidrule(r){2-4} \cmidrule(r){5-5}\cmidrule(r){6-8}
				\textrm{Variable} & $ \nu = 5 $ & $ \nu = 6 $ & $ \nu = 7 $ & $ \nu_0 $ &  $w_s$ = 0 & $w_s$ = 0.8 & $w_s$ = 1.0
				\\
				\cmidrule(r){1-8}
				\multicolumn{8}{c}{VaR}\\
				\textrm{Observations} & 455 & 455 & 455 & 455 & 455 & 455 & 455\\
				\textrm{Failures} & 7 & 8 & 11 & 159 & 224 & 15 & 8\\
				\textrm{Expected} & 4.55 & 4.55 & 4.55 & 4.55 & 4.55 & 4.55 & 4.55\\
				\textrm{Ratio} & 1.54 & 1.76 & 2.42 & 34.95 & 49.23 & 3.30 & 1.76\\
				\textrm{Missing} & 0 & 0 & 0 & 0 & 0 & 0 & 0\\
				\textrm{Traffic Light} & \textrm{green} & \textrm{yellow} & \textrm{yellow} &  \textrm{red} & \textrm{red} & \textrm{yellow} & \textrm{yellow}\\
				\textrm{Binomial Test} & \textrm{accept} & \textrm{accept} & \textrm{reject} & \textrm{reject} & \textrm{reject} & \textrm{reject} & \textrm{accept}\\		
				\cmidrule(r){1-8}
				\multicolumn{8}{c}{CVaR}\\
				\textrm{Observations} & 455 & 455 & 455 & 455 & 455 & 455 & 455\\
				\textrm{Failures} & 0 & 2 & 3 & 157 & 225 & 13 & 6\\
				\textrm{Expected} & 2.275 & 2.275 & 2.275 & 2.275 & 2.275 & 2.275 & 2.275\\
				\textrm{Ratio} & 0 & 0.88 & 1.32 & 69.01 & 98.90 & 5.71 & 2.64\\
				\textrm{Missing} & 0 & 0 & 0 & 0 & 0 & 0 & 0\\
				\textrm{Traffic Light} & \textrm{green} & \textrm{green} & \textrm{yellow} &  \textrm{red} & \textrm{red} & \textrm{yellow} & \textrm{yellow}\\
				\textrm{Binomial Test} & \textrm{accept} & \textrm{accept} & \textrm{reject} & \textrm{reject} & \textrm{reject} & \textrm{reject} & \textrm{accept}\\				
				
				\bottomrule[1.2pt]
			\end{tabular}
			\end{center}
\end{table}
\\
\begin{figure}[h!]
	\begin{center}
		\includegraphics[width=450pt]{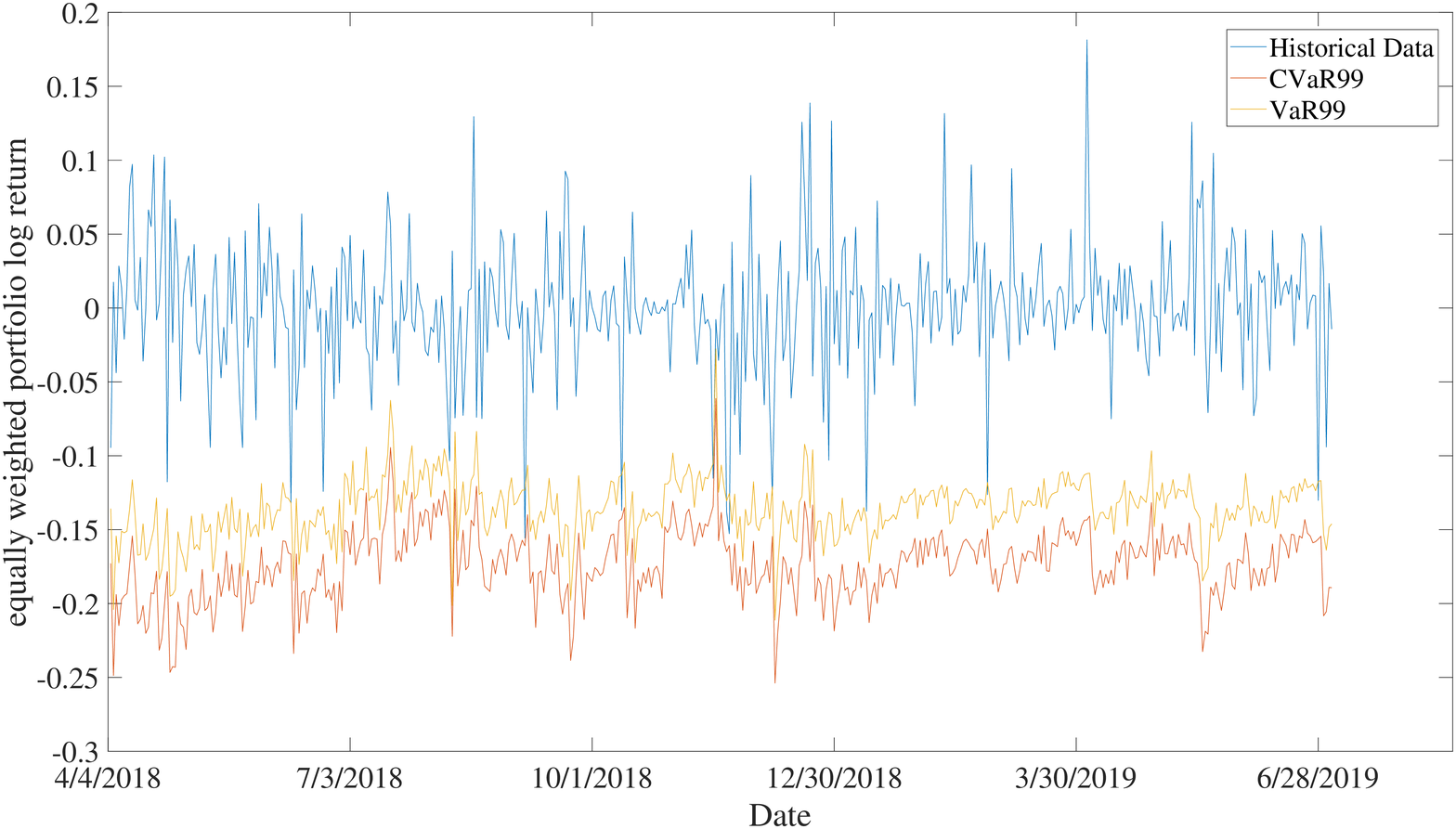}
	\caption{Backtesting Result of Multivariate {\textit t} Distribution with Five Degrees of Freedom (Gaussian Innovations)}
	\end{center}
\end{figure}
\newpage
\section{Portfolios}
\par{In this section, we perform portfolio optimization where the objective function is risk minimization. From mean-variance portfolio optimization, we have the following portfolio allocation problem with utility function $u(\cdot)$:}
\begin{equation}
	max\ E[u(\sum_{\substack{i=1}}^{d}\omega^{(i)}_{t}r^{(i)}_{t})]
\end{equation}
such that,
\begin{align}
	\sum_{\substack{i=1}}^{d}\omega^{(i)}_{t}&=1
 	\\
	L\leq \omega^{(i)}_{t} \leq U, i&=1,...,d
 \end{align}
where $L$ and $U$ are, respectively, lower and upper bounds of weights.
\subsection{Portfolio Optimization}
\par{For our portfolio optimization we use mean-variance portfolio optimization of Markowitz (1952) and CVaR portfolio optimization of Krokhmal (2002). Minimizing variance and CVaR at the 99\% confidence level are the constraints on the mean-variance and CVaR portfolios.}
\par{Using the same dataset described in Section 2.2, we perform a rolling-window optimization of 252 daily returns as the in-sample data and 455 daily returns as the out-of-sample data, while using as the benchmark the SPY with dividends.
In the empirical analysis we assume there are no transaction costs so that weights can be adjusted purely for hedging risk. Moreover, due to different trading times for crypto assets and SPY, we adjust the SPY data by adding zero returns on weekends and holidays.} 
\par{We calibrate the parameters of the ARMA(1,1)-GARCH(1,1) model (Gaussian innovation) using the in-sample data in each roll. Next, we use the multivariate {\textit t} distribution with five degrees of freedom to generate 10,000 scenarios for the one-step forecast. From the 10,000 scenarios, we find the efficient frontier, which includes the portfolios on the efficient parts of the risk-return spectrum. Then, we choose minimum risk as the criterion to construct the optimal portfolio. That is, for mean-variance portfolio optimization, we choose the point with minimum standard deviation on the efficient frontier, which we refer to as the {\itshape min Variance portfolio}. For CVaR portfolio optimization, we choose the one with minimum CVaR, which we refer to as the {\itshape min CVaR portfolio}. Corresponding weights for each crypto asset are denoted as:}
\begin{equation}
	\omega^{(i,j)}_{t},j=1,2,i=1,...,d,t=0,...,T
\end{equation}
where $j=1$ represents the {\itshape min Variance portfolio} and $j=2$ represents the {\itshape min CVaR portfolio}, $t$ is time index, and $i$ is asset index. Hence, the portfolio return $\tilde{r}^{(j)}_t$ at time $t$ is:
\begin{equation}
	\tilde{r}^{(j)}_t=\sum_{\substack{i=1}}^{d}\omega^{(i,j)}_{t}r^{(i)}_{t}
\end{equation}
Then, we move the window and repeat until the end of the out-of-sample data.
\begin{figure}[h!]
	\begin{center}
		\includegraphics[width=450pt]{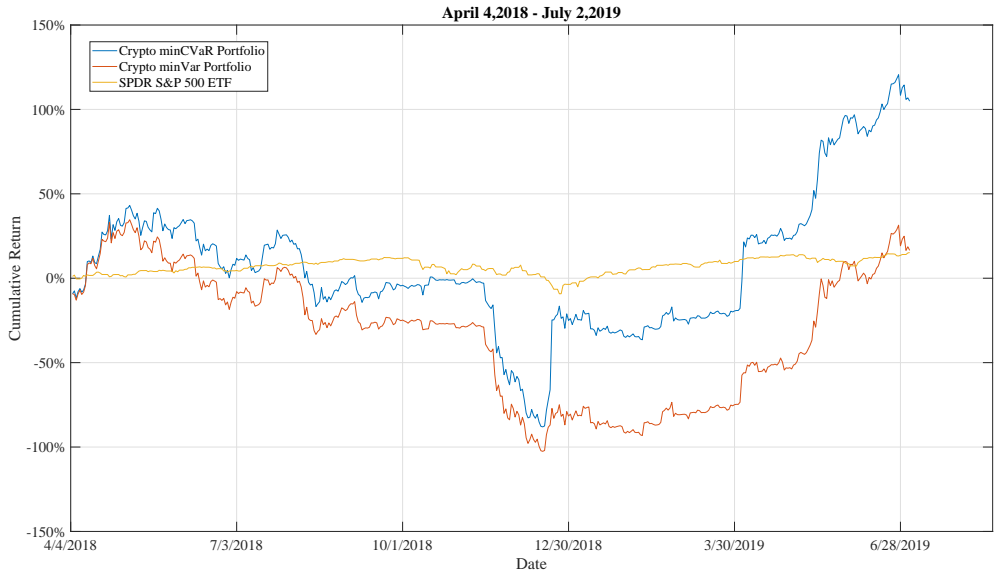}
	\caption{Horse Race of Cumulative Portfolio Return with Benchmark}
	\end{center}
\end{figure}
\par{Figure 2 shows the horse race of cumulative returns between two portfolios and the benchmark. The cumulative returns from the {\itshape min Variance portfolio} is very stable and has relatively low cumulative returns than the benchmark (SPY) all the time in the estimation window. However, the {\itshape min CVaR portfolio} outperforms the benchmark in the second quarter of 2019. The {\itshape min CVaR portfolio} has higher cumulative return than the {\itshape min Variance portfolio} all the time during the window. Our optimization results indicate that combining extremely volatile single crypto assets in a minimum-risk portfolio often outperforms a major stock market index (the S$\&$P 500).}
\par{The high return implies high risk. Hence, an analysis of risk budgeting is necessary.}

\subsection{Risk Budgeting}
\par{Two approaches based on homogeneous risk measures are applied in this study: portfolio volatility and CVaR. According to Artzner et al (1999), both measures are coherent risk measures.}
\par{Using the notation of portfolio weights $\omega^{(i,j)}_{t}$, we define $\boldsymbol\omega^{(j)}_t=(\omega^{(1,j)}_{t},...,\omega^{(d,j)}_{t})$ as the vector of weights with time index $t$ and portfolio index $j$. The risk contribution is considered under the condition of equal weights. Hence, we have $\boldsymbol\omega^{(j)}_t=(1/d,...,1/d), 1 \leq j\leq d, 0 \leq t\leq T$. To simplify, we use $\boldsymbol\omega$ and $\omega^{(i)}$ instead of $\boldsymbol\omega^{(j)}_t$ and $\omega^{(i,j)}_{t}$, respectively. And, we denote the risk measure of the $i^{th}$ asset as $RC_{i}(\boldsymbol\omega)$.}
\par{First, we focus on the volatility risk measure:}
\begin{equation}
	R(\boldsymbol\omega)=\sigma(\boldsymbol\omega)
	=\sqrt{\boldsymbol\omega^{T}\Sigma\boldsymbol\omega}	
\end{equation}
where $\Sigma$ is covariance matrix between asset returns. The marginal risk and risk contribution of the $i^{th}$ crypto asset are:
\begin{align}
	\frac{\partial R(\boldsymbol\omega)}{\partial \omega^{(i)}}&=
	\frac{(\Sigma x)_{i}}{\sqrt{\boldsymbol\omega^{T}\Sigma\boldsymbol\omega}}
\end{align}
\begin{align}
	RC^{Vol}=RC_{i}(\boldsymbol\omega)&=\omega^{(i)} \frac{(\Sigma x)_{i}}{\sqrt{\boldsymbol\omega^{T}\Sigma\boldsymbol\omega}}
\end{align}
\par{The second risk measure is CVaR at the confidence level $\alpha \in (0,1)$. As defined in equations (11) and (12), we let $RC^{VaR}=CVaR_{\alpha}$.
In the risk budgeting analysis, we set $\alpha=0.01$ and keep the same setting for the in-sample and out-of-sample data as in Section 2.2.}
\subsubsection{In-sample and Out-of-sample Risk Budgeting}
\par{Portfolio volatility and CVaR are used as the risk measures for the in-sample data. According to the results reported in \:Table 3, some similarities can be found between $RC^{Vol}$ and $RC^{VaR}$. Binance coin (BNB), bitcoin cash (BCH), and EOS (EOS) seem to have relatively higher risk than the other four crypto assets. Meanwhile, bitcoin has the lowest risk contribution in both cases.  Later, we check our conclusion with the out-of-sample data. Figures 3 and 4 show that bitcoin is the risk diversifier and EOS is the risk contributor of the equally weighted portfolio.\footnote{The risk diversifier is a diversification tool which yields higher long-term returns and lower risk in a portfolio, while the risk contributor is the opposite.} Due to the larger number of observations involved in the out-of-sample analysis, which implies more accuracy, we conclude the risk diversifier is bitcoin and the risk contributor is EOS.}
\\
   \begin{table}[h!]
	\begin{center}		
		\label{tab:Table 3}
		\caption{In-Sample Risk Budgeting in 07/25/2017-04/03/2018}
			\begin{tabular}{l l l l l l l l}
				\toprule[1.2pt]
				\textrm{Method}&{BTC}&{ETH}&{XRP}&{LTC}&{BCH}&{EOS}&{BNB}\\
				\cmidrule(r){1-8}
				\textrm{$RC^{Vol}$}&0.0059&0.0073&0.0088&0.0089&0.0103&0.0106&0.0124\\
				\textrm{$RC^{Vol}$(\%)}&9.22&11.39&13.67&13.90&16.00&16.47&19.34\\
				\\
				\textrm{$RC^{VaR}$}&0.028&0.0331&0.0479&0.0434&0.0545&0.049&0.0536\\
				\textrm{$RC^{VaR}$(\%)}&9.05&10.68&15.49&14.01&17.59&15.85&17.33\\
				\bottomrule[1.2pt]
			\end{tabular}
			\end{center}
\end{table}
\begin{figure}[h!]
	\begin{center}
		\includegraphics[width=450pt]{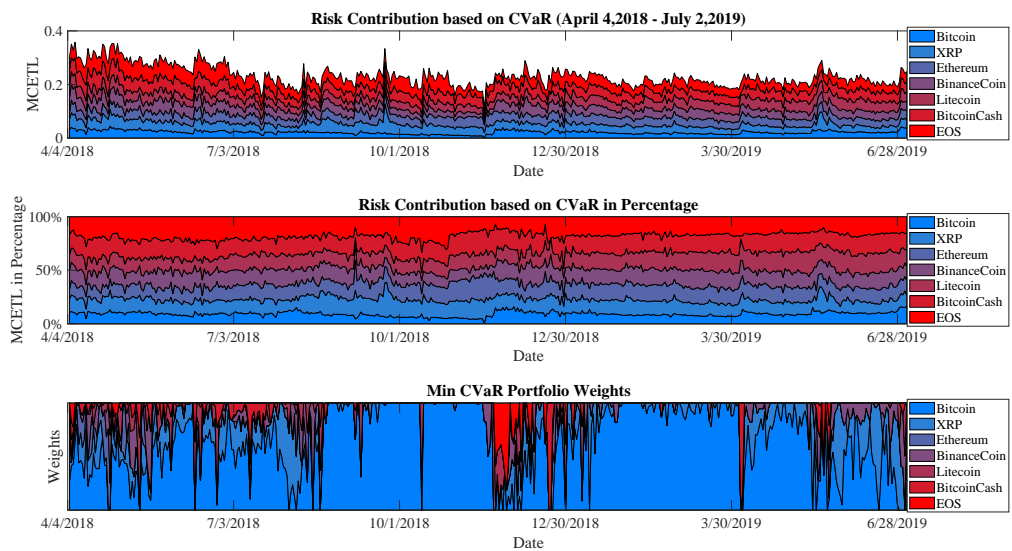}
	\caption{Out-of-Sample Risk Budgeting based on CVaR}
	\end{center}
\end{figure}
\begin{figure}[h!]
	\begin{center}
		\includegraphics[width=450pt]{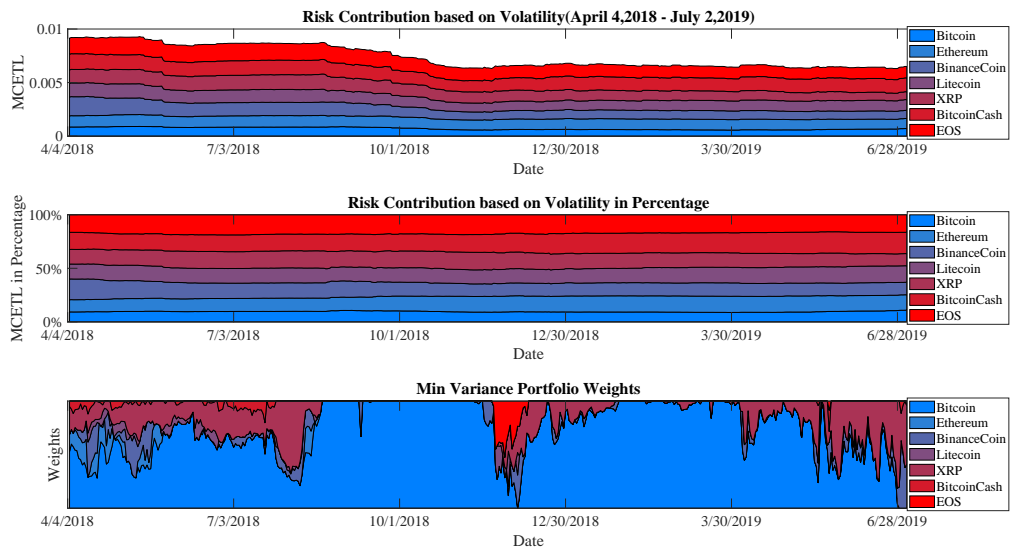}
	\caption{Out-of-Sample Risk Budgeting based on Volatility}
	\end{center}
\end{figure}
\subsubsection{Measures of Risk-Adjusted Return}
\par{Risk-adjusted returns allow investors to assess high-risk versus low-risk investments. Here we analyze four measures of risk-adjusted return:}
\begin{itemize}
	\item [1.]
	The Maximum Drawdown (MDD) is defined as the maximum loss incurred from peak to bottom during a specified 	period of time $[0,T]$.
	\begin{equation}
		MDD(T)=sup_{t\in [0,T]}[sup_{s \in [0,t]}(S_{s}-S_{t})]
	\end{equation} 
	\item [2.]
	The Sharpe ratio (see Sharpe, 1994) is defined as
	\begin{equation}
		Sharpe(T)=\frac{\bar{R}(T)-R_f}{\sigma_P}
	\end{equation}
	where $\bar{R}(T)=\frac{1}{T}\sum_{t=0}^{T}\tilde{r}_t$ is the mean portfolio return, $R_f$ is the risk-free rate, and $\sigma_P$ is the portfolio volatility within $[0,T]$.
	\item [3.]
	The M2 ratio (see Modigliani and Modigliani, 1997) is derived from the Sharpe ratio with defined benchmark 	volatility as $\sigma_M$.
	\begin{align}
		M2(T)&=Sharpe(T)\sigma_M+R_f
	\end{align}
	\item [4.]
	The Rachev ratio (see Rachev et al, 2008) is defined as the ratio between the CVaR of opposite of the excess return at the $1-\alpha$ confidence level and the CVaR of the excess return at the $1-\beta$ confidence level.
	\begin{equation}
		Rachev_{\alpha,\beta}(T)=\frac{CVaR_{\beta}(R_f-R_P)}{CVaR_{\alpha}(R_P-R_f)}
	\end{equation}
	In our analysis, we set $\alpha=\beta=0.01$.
	\end{itemize}
\par{The results, reported in Table 4, are based on the out-of-sample data. In our calculation, we use the US 10-year Treasury yield curve rates as the risk-free rate. Maximum Drawdown (MDD) for the two optimal portfolios are very close, which are much higher than the SPY. Sharpe ratios indicate that the {\itshape min CVaR portfolio} outperforms the SPY. M2 ratios and Rachev ratios ($\alpha=\beta=0.01$) of the {\itshape min Variance portfolio} and the SPY are almost identical. Based on the four ratios analyzed, we conclude that the {\itshape min CVaR portfolio} has the highest risk-adjusted return and the {\itshape min Variance portfolio} and the benchmark SPY have similar performance.}
\\
\begin{table}[h!]
	\begin{center}		
		\label{tab:Table 4}
		\caption{Measures of Risk-Adjusted Return Based on Out-of-Sample Data in 04/04/2018-07/02/2019}
			\begin{tabular}{l c c l}
				\toprule[1.2pt]
				\textrm{Measures}&{min CVaR portfolio}&{min Variance portfolio}&{SPY}\\
				\hline
				\textrm{MDD}& 0.7307 & 0.7464 & 0.1935\\
				\textrm{Sharpe ratio}& 0.0502 & 0.0078 & 0.0333\\
				\textrm{M2 ratio}& 0.0023 & 0.0004 & 0.0003\\
				\textrm{Rachev ratio}& 1.7588 & 1.0124 & 1.0246\\
				\bottomrule[1.2pt]
			\end{tabular}
			\end{center}
\end{table}
\section{Option Pricing}
\par{If options on a crypto asset index existed, standard methodologies for hedging would exist (protective put  buying to create a nonlinear payoff profile).  However, in the absence of options where the underlying is a crypto asset index, hedging is still possible if there is a crypto asset index. 
The dynamic trading strategy to do so is a strategy formulated in the 1980s by the advisory firm of Leland O'Brien Rubinstein Associates and called “portfolio insurance”. 
We apply this strategy to calculate the fair value of crypto options in which the underlying is a crypto asset index. 
The option valuation model is applied to the optimal crypto asset portfolio derived in Section 3 (i.e., the minimum-risk crypto portfolio) in order to reduce price risk by hedging.}
\par{We start by finding an appropriate distribution for the innovations for the portfolio returns.
Similar to Section 2, we keep the GARCH-type model for returns as given by equations (2)-(5) because the ARMA effect is irrelevant for risk-neutral option pricing. Then, we fit the model to the generalized hyperbolic distribution introduced by Barndorff-Nielso (1977) to find the appropriate distribution for the innovation $\epsilon$. After carefully backtesting, the results indicate that the innovations from both the {\itshape min variance portfolio} returns and {\itshape min CVaR portfolio} returns follow the normal inverse Gaussian distribution (NIG($\alpha,\beta,\delta,\mu$)) with the moment-generating function:
\begin{equation}
	M(z)=e^{\mu z +\delta \sqrt{\alpha^2-\beta^2}-\delta \sqrt{\alpha^2-(\beta+z)^2}}
\end{equation}
where $\alpha$ is the tail parameter, $\beta$ is the asymmetry parameter, $\delta$ is the scale parameter, and $\mu$ is the location parameter. 
Therefore, we offer a reliable model: the ARMA(1,1)-GARCH(1,1) model with NIG innovation for the minimum-risk portfolio returns. Then, we obtain the valuation of crypto options by passing the natural world to the unique equivalent martingale measure via the Esscher transform (see Gerber and Shiu, 1994). 
Following Chorro (2012), we apply the following methodology to deal with the NIG innovation case:}
\begin{itemize}
	\item[1.]
	Select the sample: 455 daily portfolio returns (i.e., the {\itshape min Variance portfolio} returns).
	\item[2.]
	Estimate model (2)-(5) with NIG innovations based on the selected sample.
	\item[3.]
	Set $t=0$, fit innovations with NIG($\alpha,\beta,\delta,\mu$) and output parameters and one-step 	forecast of conditional variance $\sigma_1^2$.
	\item[4.]
	Repeat the following (a)-(d) steps for $t=1,...,T$. We set maturity $T$ at six months and use the 6-month Treasury yield curve rates as the risk-free rate $r$.
	\begin{itemize}
		\item[(a)]
		Solve the following function to get $\theta_t$.
		\begin{equation}
			\sqrt{\alpha^2-(\beta+\theta)^2}-\sqrt{\alpha^2-(\beta+1+\theta)^2}=\frac{r-\mu}{\delta}
		\end{equation}
		\item[(b)]
		Renew $\beta$ as $\beta+\sqrt{\sigma_{t}}\theta_{t}$.
		\item[(c)]
		Generate $\epsilon_{t+1}$ from NIG($\alpha,\beta,\delta,\mu$).
 		\item[(d)]
		Compute $r_{t+1}$ and $\sigma_{t+2}$.
	\end{itemize}
	\item[5.]
	Setting \$100 as the initial capital, the future value at $T$ is
	\begin{equation}
		S_T=100e^{\sum_{\substack{t=1}}^{T}r_t}
	\end{equation}
	\item[6.]
	Repeat steps 4 and 5 for $N=10,000$ times as suggested by Chorro (2012).
	\item[7.]
	The approximated call option price and put option price are
	\\
	\begin{align}
		\hat C(t,K;T)&=e^{-r(T-t)}\frac{1}{N}\sum_{\substack{i=1}}^{N}max(S_{T,i}-K,0)
		\\
		\hat P(t,K;T)&=e^{-r(T-t)}\frac{1}{N}\sum_{\substack{i=1}}^{N}max(K-S_{T,i},0)
	\end{align}
	where $K$ is the exercise strike.
\end{itemize}
\par{ Using the enhanced Monte Carlo method, we calculate the fair price of crypto options. We capture the short-term behavior of the option smile found for equity options based on the {\itshape min Variance portfolio} in Figure 5(a). The “smile” is well illustrated in Figure 5(c). As can be seen in Figure 5(b), the curvature of the option smile increases as time to maturity $T$ decreases. For the options based on the {\itshape min CVaR portfolio} returns, we have similar figures as shown in Figure 6.}
\par{Note that our method for the valuation of crypto options can be applied to all European style options. 
Options give the option buyer the right but not the obligation to purchase or sell the underlying asset at a predetermined price. 
With the put options given by equation (29), an investor can hedge the risk for the minimum-risk crypto portfolios with a specified strike price. 
We cannot, of course, compare our results with option prices in the crypto market because such market prices do not exist. 
Yet, we can claim that if such options on crypto assets are introduced, they should follow closely our theoretical prices after adjusting for market frictions and design feature nuances.}
\begin{figure}
  \begin{subfigure}{500pt}
    \centering\includegraphics[width=500pt]{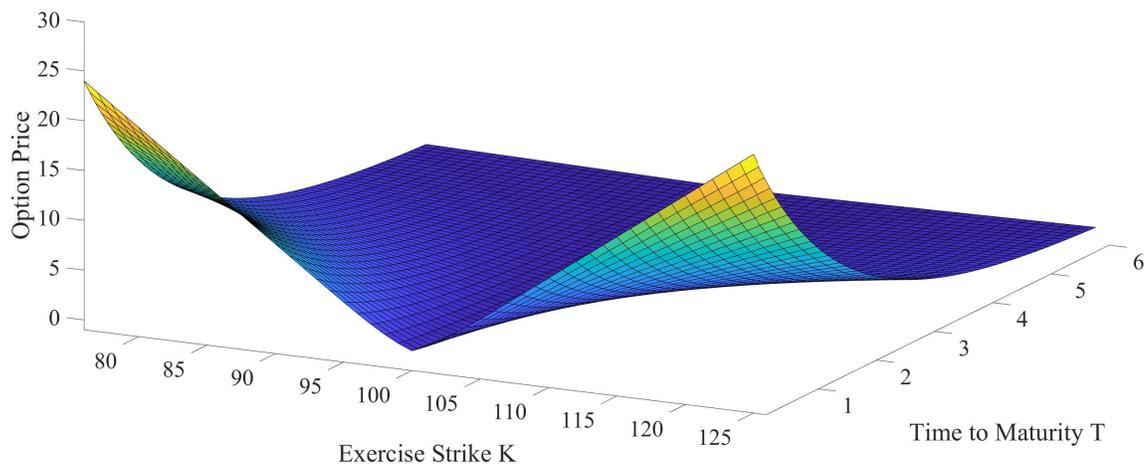}
    \caption{}
  \end{subfigure}
  
  \begin{subfigure}{250pt}
    \centering\includegraphics[width=250pt]{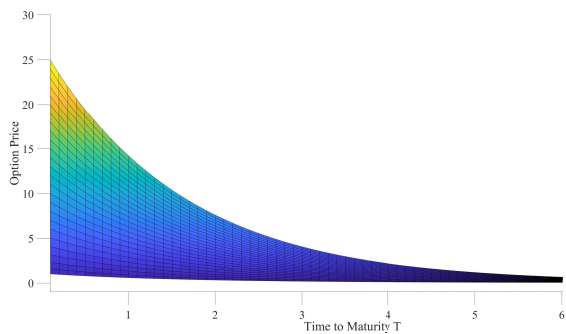}
    \caption{}
  \end{subfigure}
  \begin{subfigure}{250pt}
    \centering\includegraphics[width=250pt]{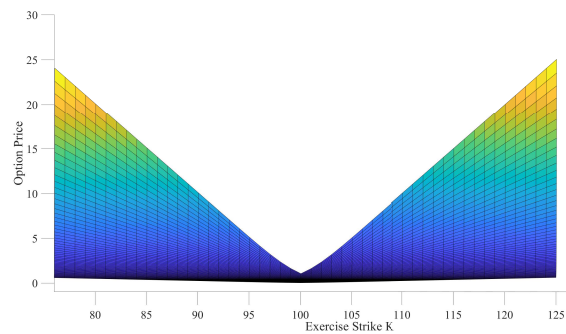}
    \caption{}
  \end{subfigure}
\caption{Option Prices based on the {\itshape min CVaR portfolio}}
\end{figure}
\begin{figure}
  \begin{subfigure}{500pt}
    \centering\includegraphics[width=500pt]{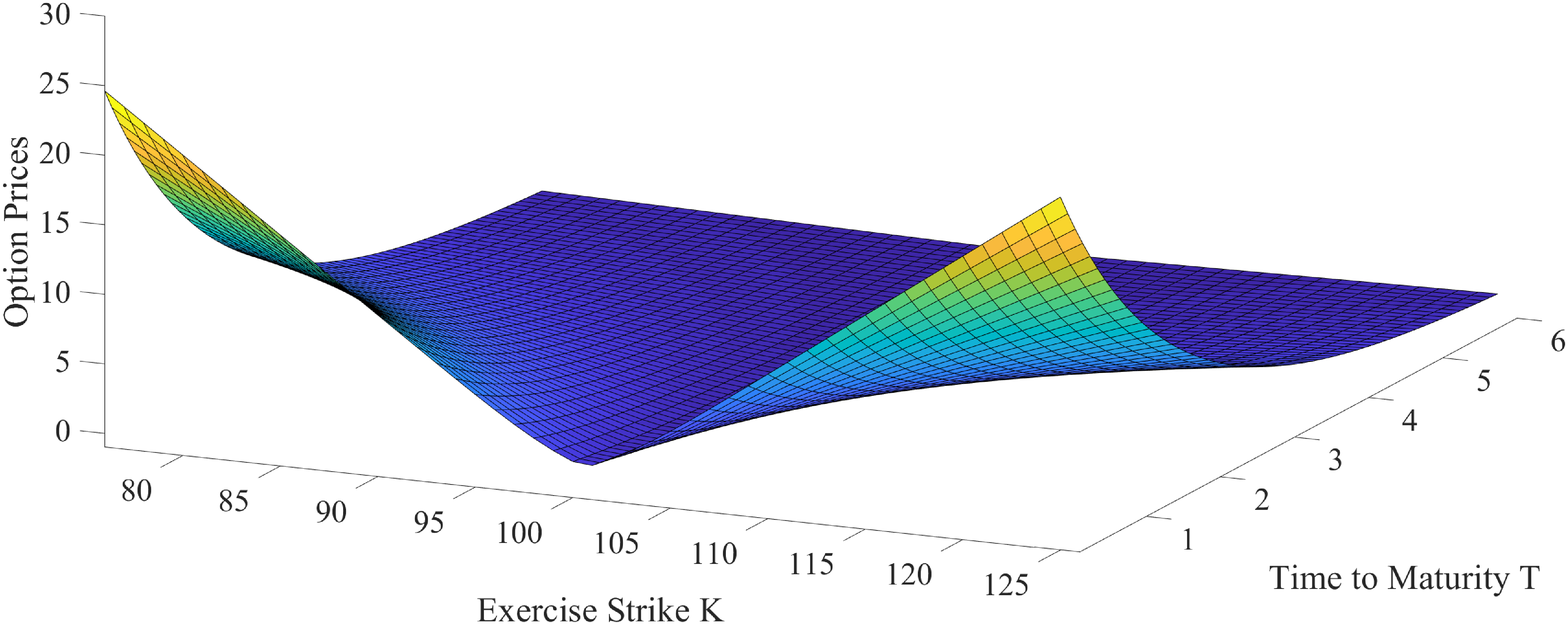}
    \caption{}
  \end{subfigure}
  
  \begin{subfigure}{250pt}
    \centering\includegraphics[width=250pt]{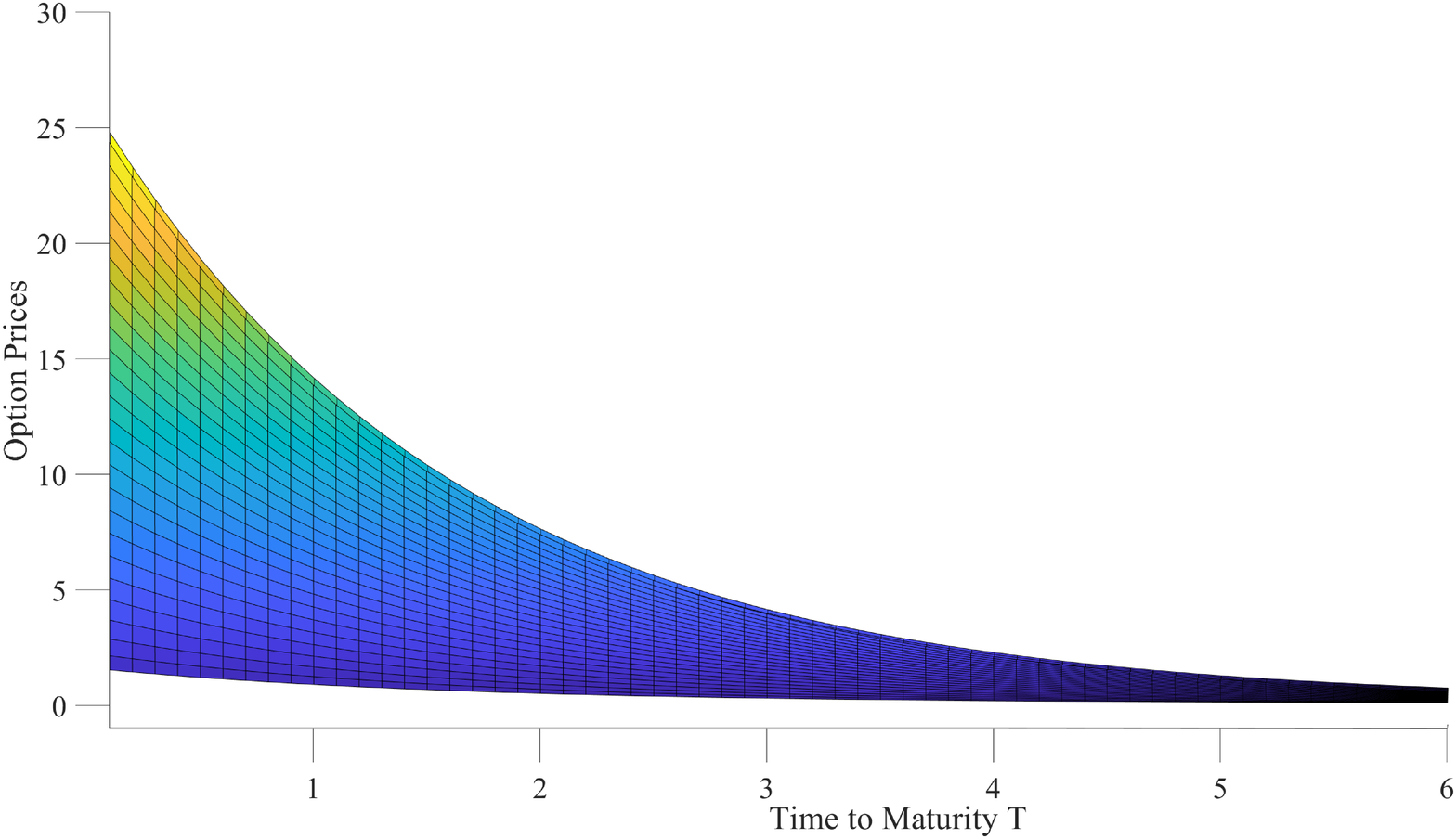}
    \caption{}
  \end{subfigure}
  \begin{subfigure}{250pt}
    \centering\includegraphics[width=250pt]{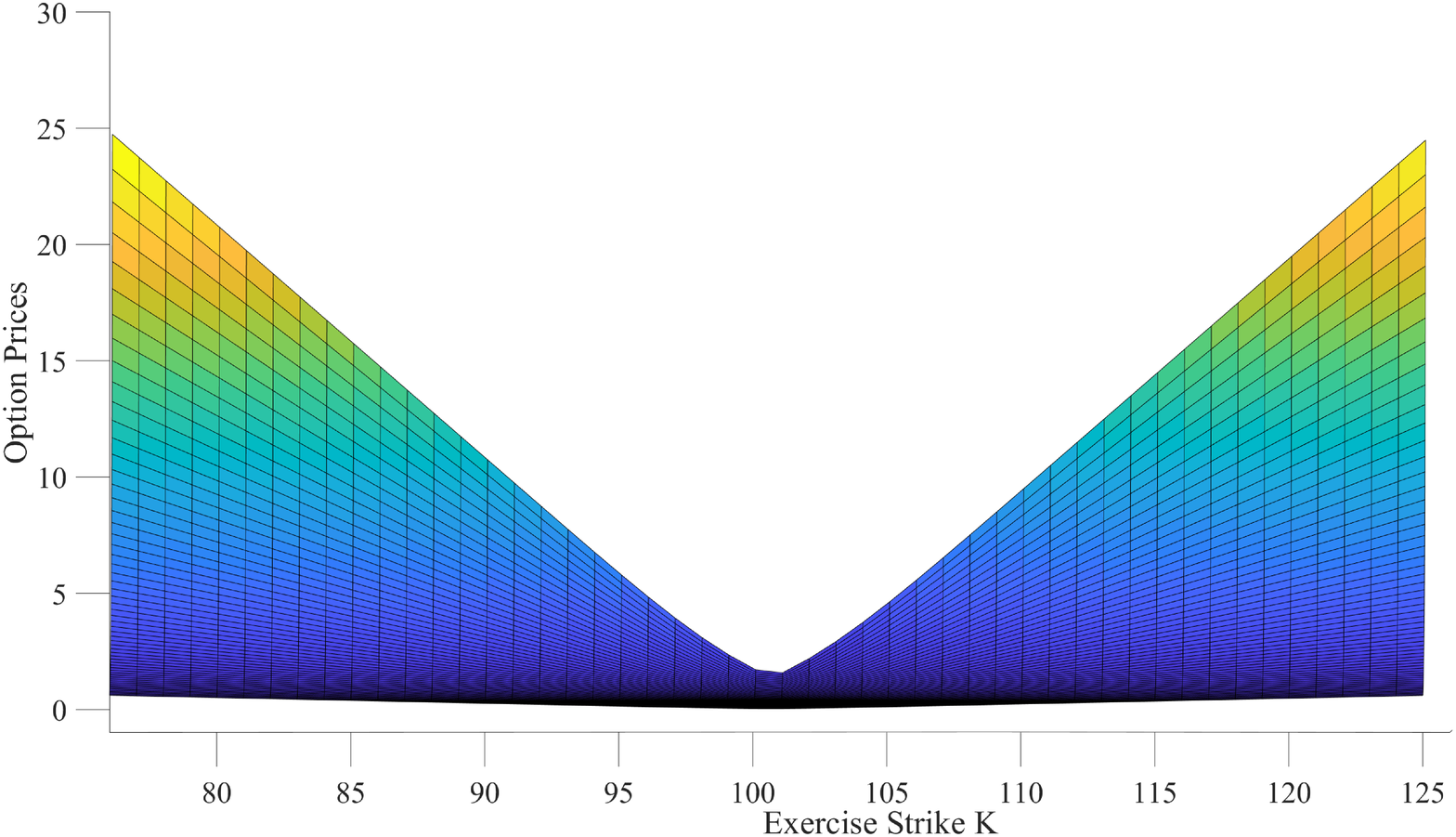}
    \caption{}
  \end{subfigure}
\caption{Option Prices based on the {\itshape min CVaR portfolio}}
\end{figure}
\section{Conclusion}
\par{In this paper, we conduct an empirical analysis of the minimum-risk portfolios of major crypto assets. We start by fitting the sample returns using the ARMA(1,1)-GARCH(1,1) model with different multivariate distributions. Using VaR and CVaR as measures for backtesting, we conclude that the optimal model is ARMA(1,1)-GARCH(1,1) (Gaussian innovation) with multivariate {\textit t} distribution with five degrees of freedom. Based on the model, we propose the {\itshape min Variance portfolio} and the {\itshape min CVaR portfolio} using mean-variance and CVaR portfolio optimization, respectively. The first major contribution of this paper is demonstrating that the combination of extremely volatile single crypto assets in a minimum-risk portfolio often outperforms a major  stock market index, the S$\&$P 500. With risk budgeting, we find the risk diversifier is bitcoin and the risk contributor is EOS. Also, based on four measures of risk-adjusted return, we conclude that {\itshape min CVaR portfolio} has the highest risk-adjusted return while the {\itshape min Variance portfolio} and the benchmark (the S$\&$P 500) have similar performance. We then offer a novel solution for calculating the fair value of crypto options based on two minimum-risk portfolios for hedging. 
If such options on crypto asset indexes are introduced, we would expect that they should follow closely our theoretical prices after adjusting for market frictions and design feature nuances.}
\nocite{tsay2010financial}
\nocite{campbell1997econometrics}
\nocite{WhittleARMA}
\nocite{Demarta2005}
\nocite{hitaj2013portfolio}
\nocite{wang2009}
\nocite{Markowitz1952}
\nocite{Krokhmal2002}
\nocite{Artzner1999}
\nocite{BarndorffNielson1977}
\nocite{chorro2012option}
\nocite{Duffie2010DynamicPricing}
\nocite{8thoptionsfutures}
\nocite{QuantitativeFinance}
\nocite{hurst1999option}
\nocite{chan2017statistical}
\nocite{chu2017garch}
\nocite{bruder2012managing}
\nocite{sharpe1994sharpe}
\nocite{stoyanov2007optimal}
\nocite{RobbieSusan2018}
\nocite{borak2005}
\nocite{Modigliani1997}
\nocite{Rachev2008}
\nocite{Gerber1994}
\nocite{Thakor2019}
\nocite{Bollerslev1986}
\nocite{Balcilar2017}
\bibliography{references}
\bibliographystyle{apalike}

\end{document}